\documentclass[twocolumn,secnumarabic,amssymb, nobibnotes, superscriptaddress, aps, pre]{revtex4-1}
\usepackage{amsmath}
\usepackage{graphicx}
\usepackage[dvipsnames]{xcolor}

\begin{document}

\title{Laser-driven pointed acceleration of electrons with preformed plasma lens}%

\author{K.A. Ivanov}
\email{akvonavi@gmail.com}
\affiliation{Faculty of Physics, Lomonosov Moscow State University, 119991, Moscow, Russia \looseness=-1}
\affiliation{Lebedev Physical Institute of Russian Academy of Sciences, 119991, Moscow, Russia \looseness=-1}

\author{D.A. Gorlova}
\affiliation{Faculty of Physics, Lomonosov Moscow State University, 119991, Moscow, Russia \looseness=-1}
\affiliation{Institute for Nuclear Research of Russian Academy of Sciences, 117312, Moscow, Russia \looseness=-1}

\author{I.N. Tsymbalov}
\affiliation{Faculty of Physics, Lomonosov Moscow State University, 119991, Moscow, Russia \looseness=-1}
\affiliation{Institute for Nuclear Research of Russian Academy of Sciences, 117312, Moscow, Russia \looseness=-1}

\author{I.P. Tsygvintsev}
\affiliation{Keldysh Institute of Applied Mathematics of Russian Academy of Sciences, 125047 Moscow, Russia \looseness=-1}

\author{S.A. Shulyapov}
\affiliation{Faculty of Physics, Lomonosov Moscow State University, 119991, Moscow, Russia \looseness=-1}

\author{R.V. Volkov}
\affiliation{Faculty of Physics, Lomonosov Moscow State University, 119991, Moscow, Russia \looseness=-1}

\author{A. Savel'ev}
\affiliation{Faculty of Physics, Lomonosov Moscow State University, 119991, Moscow, Russia \looseness=-1}
\affiliation{Lebedev Physical Institute of Russian Academy of Sciences, 119991, Moscow, Russia \looseness=-1}

\date{October 2023}%

\begin{abstract}

The simultaneous laser-driven acceleration and angular manipulation of the fast electron beam is experimentally demonstrated. The bunch of multi-MeV energy charged particles is generated during the propagation of the femtosecond laser pulse through the near-critical plasma slab accompanied by plasma channeling. Plasma is formed by the controlled breakdown of a thin-tape target by a powerful nanosecond prepulse. The electron beam pointing approach is based on the refraction of a laser pulse in the presence of a strong radial density gradient in the breakdown of the tape with a small displacement of the femtosecond laser beam relative to the breakdown symmetry axis. A shift of several micrometers makes it possible to achieve beam deflection by an angle up to 10 degrees with acceptable beam charge and spectrum conservation. This opens up opportunities for in-situ applications for scanning objects with an electron beam and the multistage electron beam energy gain in consecutive laser accelerators without bulk magnetic optics for particles. Experimental findings are supported by numerical Particle-In-Cell calculations of laser-plasma acceleration and hydrodynamic simulations.

\end{abstract}
\maketitle

\section{\label{sec:level1}Introduction}

The generation of energetic collimated electron beams with a high charge using laser-plasma accelerators has become one of the main area of application for terawatt (TW) and petawatt (PW) laser systems \cite{1Malka,2Joshi,3Ledigham}. Thanks to Laser Wakefield Acceleration (LWFA \cite{4Kurz}) and Direct Laser Acceleration (DLA \cite{5Hussein,6Arefiev}) in relativistic plasma channel, together with the rapid development of high repetition rate TW-class lasers the new phenomena related to astrophysics \cite{7Takabe}, nuclear photonics \cite{8Nedorezov}, biomedicine and diagnostics \cite{9Nakajima,10Labate}, etc. has become available to a wide community of researchers literally within the laboratory. The latter circumstance imposes fairly severe restrictions on the size of laboratory sources utilizing compact laser systems with peak power of up to 100 TW (including new-generation systems with peak power of a few TW with a repetition rate up to a kHz level \cite{11Antipenkov}). Besides, the design of the electron beam generator itself (target assembly and adjacent elements) must have reasonable dimensions and provide simple and stable operation as well as control over the beam. One should also consider the bunch acceleration efficiency in terms of mean energy and charge. Instead of a single accelerating stage by an extremely powerful laser pulse the use of a few laser pulses can provide consecutive bunch energy gain in a number of acceleration stages driven by less-demanding and modest laser systems \cite{12Kim,13Zhu,14Jin}.

For various application scenarios the angular targeting and pointing of the electron beam are of key importance, whether it concerns the optimal injection into the subsequent stage of acceleration \cite{15Massimo,16Starodubyseva}, injection schemes \cite{17Khudiakov} relevant to the AWAKE experiment \cite{18Moschuering}, scanning of a sample, directing the bunch into different detectors, radio medicine applications including FLASH therapy \cite{19Favaudon}, etc. For this, standard accelerator technology can be used: magnetic lenses and dipoles. But their key drawback is high energy selectivity, whereas the laser acceleration of particles beam with high charge leads to rather wide energy spectrum \cite{20Gahn}. Also worth noting the bulkiness and slow adjustment. Therefore, the issue of controlling the beam pointing is of great importance not only from the point of view of compactness, but also the preservation of a high charge. For these purposes, several purely plasma methods have been proposed for the LWFA scheme. Among them, we can single out the waveguide propagation of a pulse, when a laser beam is introduced onto the axis of an electron bunch through a curved plasma channel \cite{21Nakajima,22Luo}. It was also proposed to use a pulse with an inclined wave front, which will turn in a low-density plasma and take the electron beam behind it in the Wakefield acceleration mode \cite{23Mittelberger,24Zhu}.

In this work, we experimentally demonstrate the possibility of controlling the ejection angle of an electron bunch with particle energy exceeding a few MeV accelerated by a 1 TW laser pulse in a plasma channel. The latter is formed in a plasma sheet of near critical density when a thin (15 $\mu$m) mylar tape is hole bored by an additional nanosecond prepulse (100 mJ, 8 ns, $10^{13}$ W cm$^{-2}$), coming ahead of the main accelerating femtosecond pulse (50 mJ, 50 fs, $5 \times 10^{18}$ W cm$^{-2}$) by a few nanoseconds. The prepulse forms a breakdown with strong radial density gradient. Precise adjustment of the breakdown axis with respect to the femtosecond pulse axis makes it possible to control the trajectory of the relativistic plasma channel due to laser beam refraction. The accelerated electron beam preserves its characteristics (energy, spectrum, divergence) at a high degree. The findings are confirmed by simulations and analytical model. 

\section{Main experimental results}%

\begin{figure}[h]
\includegraphics[width=0.9\linewidth]{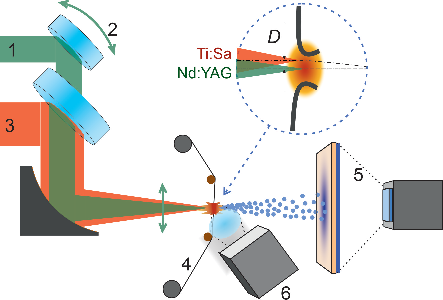}
\caption{\label{Figure1} Experimental scheme: 1 - Nd:YAG nanosecond prepulse, 2 - adjustable mirror for Nd:YAG radiation, 3 - femtosecond pulse, 4 - tape target, 5 - electron beam detector (Lanex scintillation screen and camera with lens), 6 - camera with microobjective collecting the scattered light from plasma (perpendicular to observer), \textit{D} - deviation of the Ti:Sa laser pulse axis relative to the Nd:YAG axis (target breakdown symmetry axis).}
\end{figure}

Previously, we proposed an efficient laser-driven DLA electron beam source using subcritical plasma and moderate peak power (1 TW) laser system \cite{25Ivan} operating at 10 Hz. The electron bunch had a $\sim$50 pC charge for the energy range above $\sim$3 MeV. To form a plasma with the required electron density longitudinal profile, we used a nanosecond prepulse from a Nd:YAG laser at a wavelength of 1064 nm, focused to a peak intensity $\sim10^{12}$ W cm$^{-2}$ onto the surface of thin plastic tape target at an angle close to the normal. After the peak of the prepulse with delay of several nanoseconds the terawatt radiation from the TiSa laser system (805 nm, 50 fs, $>10^{18}$ W cm$^{-2}$) was focused into the preformed plasma along the same optical axis. With such an interaction geometry, efficient generation of an electron beam is possible under Direct Laser Acceleration in the self-focusing induced plasma channel. The optimal delay between the peaks of the pulses depends on the target thickness and prepulse intensity, since it determines nanosecond plasma evolution dynamics of the tape target. In the current experimental campaign we used a 15 $\mu$m thick mylar tape target. The prepulse intensity was on the level of $10^{13}$ W cm$^{-2}$ with a focal diameter 13 $\mu$m spot at full width at half maximum (FWHM), which resulted in the optimal delay where the electron beam is generated equal to $\sim$2 nanoseconds. The femtosecond pulse was focused into a 4.0$\pm$0.2 $\mu$m FWHM spot to a peak intensity $\approx5\times10^{18}$ W cm$^{-2}$. The relative amplitude of the amplified spontaneous emission (ASE) pedestal in the domain 10-100 picoseconds prior to main pulse was on the level of 10$^{-8}$.
Obviously, the density distribution created by the nanosecond pulse is radially dependent and the plasma acts as a lens: one can estimate the laser beam refraction under its propagation with a small deviation \textit{D} from the longitudinal symmetry axis of the plasma hole in the tape target. For this we implemented the updated experimental setup (Fig.\ref{Figure1}) which makes possible to shift the axis of the heating nanosecond pulse with respect to the femtosecond pulse axis by tilting the mirror (2) in the Nd:YAG laser path.

\begin{figure}[h]
\includegraphics[width=1\linewidth]{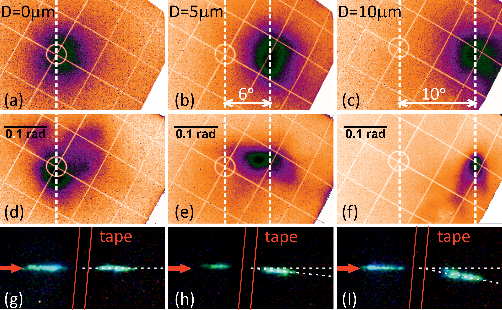}
\caption{\label{Figure2} Upper row - the averaged image of the electron beam on the detector with \textit{D} = 0 (a), 5 (b) and 10 (c) $\mu$m. Second row - the electron beam stamp in single shots at \textit{D} = 0 (a), 5 (b) and 10 (c) $\mu$m. Bottom row - the plasma scattered light signal in the optical range under the same femtosecond pulse propagation conditions: \textit{D} = 0 (coaxial, g), 5 $\mu$m (h) and 10 $\mu$m (i). Red arrow indicates the incident laser beam direction. The red lines mark the tape initial position.}
\end{figure}

The alignment of the axes of nano- and femtosecond pulses was controlled by transferring images of the waist of the off-axis parabolic mirror by a microobjective to a CCD camera with high optical resolution. The discreteness of the \textit{D} was determined by a stepper actuator on the mount of the Nd:YAG laser mirror and was 2.5$\pm$0.5 $\mu$m. The electron beam was detected on a Lanex scintillating screen in the direction of laser radiation. A metal filter (300 $\mu$m of tungsten) directly in front of the scintillator cut off electrons with energies below $\sim$3 MeV without strong additional beam divergence after scattering in the filter. Scattered light from the plasma in the visible range was also detected in the perpendicular to laser propagation plane as the femtosecond pulse passed through the tape. For this, the region of interest was transferred to a CCD camera using a microobjective.

Fig.\ref{Figure2} (top row) illustrates averaged over 100 shots images of the electron beam on the Lanex screen for coaxial propagation of femtosecond and nanosecond pulses (a), and for \textit{D} = 5 and 10 $\mu$m, (b) and (c) correspondingly. Of course, from shot to shot, the position of the beam varies somewhat across the detector, which is due to the instabilities of the transverse intensity distribution across the focal spot of the nanosecond laser, the temporal jitter of the Q-switch, which ultimately affect the breakdown parameters in the tape. In the selected single shots (see second row in Fig.\ref{Figure2}), the beam divergence does not exceed 0.1 rad. The maximum absolute electron beam displacement on the Lanex corresponds to an angle of deflection from the original axis by $\sim$10 degrees, Fig.\ref{Figure2}(c). 

No evident decrement of average scintillation screen luminosity was observed experimentally, which suggests that the beam charge did not vary significantly. The image of the plasma channel scattered light follows the electron beam as seen in the third row in Fig.\ref{Figure2}. It is worth noting that the absolute position of the tape relative to the waist of the femtosecond beam did not require adjustment in the case of a transverse displacement of the breakdown axis. This greatly simplifies the handling of electron beam for applications.

\section{Numerical simulations and discussion}

\begin{figure}[h]
\includegraphics[width=0.9\linewidth]{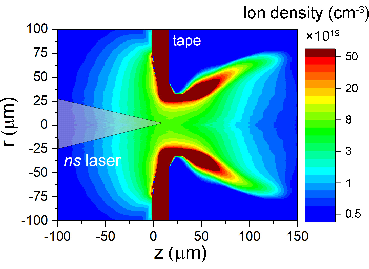}
\caption{\label{Figure3} Calculated distributions (r,z) of the ions concentration in plasma of the PET tape.}
\end{figure}

The deflection angle  of the laser beam $\theta$ during propagation along the Oz axis is determined by the change in the refractive index $n$ along the radial coordinate: $\dfrac{d\theta}{dz}=n^{-1}$$\dfrac{dn}{dr}$. Whereas, according to the Drude model, the refractive index depends on the electron density $n_e$ in plasma. The preformed plasma with radial density gradient acts as a lens, leading to the refraction of the femtosecond pulse. Additional modulation into the local refractive index (which causes, in particular, the formation of a plasma channel) is introduced by field ionization and the ponderomotive action of light during the relativistic laser pulse propagation through the plasma. We made hydrodynamic simulations of the tape ablation by the nanosecond laser pulse. Then PIC simulations of the electron acceleration process using the simulated plasma profile were performed. PIC simulations were made for different deviations \textit{D} from the plasma longitudinal axis.
The target ablation and boring was modeled in hydrodynamic software package 3DLINE \cite{26Krukovskiy}. The one-temperature one-fluid model of a quasi-neutral plasma was used. The local ionization multiplicity was determined from the stationary distribution obtained in the framework of collisional-radiative equilibrium using the THERMOS code \cite{27Vichev}. This approximation works well on the considered scale of characteristic times ($\sim$ns) and laser intensities ($10^{12}-10^{13}$ W cm$^{-2}$), and detailed consideration of the ionization kinetics is not required. The equation of state was calculated using the Thomas-Fermi model with semi-empirical corrections in the phase transition region using the FEOS code \cite{28Faik}. The target material was PET (initial density 1.38 g/cm${^3}$, temperature 270 K), thickness 15 $\mu$m. The laser parameters where similar to the experiment: 1064 nm, 8 ns pulse duration FWHM, peak intensity $10^{13}$ W cm$^{-2}$. The simulation was carried out in two-dimensional axially symmetric (r,z) geometry on a fixed grid with a size of 0.5x0.945 mm. Fig.\ref{Figure3} illustrates the resulting distribution of ions concentration with strong radial dependence. The data for the optimal delay was then integrated into the PIC code SMILEI \cite{29Derouillat}.
We first carried out 3D modeling to ensure the accurate reproduction of the experimental conditions in the case of co-axial propagation of the two pulses. The target initially consisted of neutral carbon. To save computational time and resources the simulations with shifted propagation from the symmetry axis were performed in simplified 2D3V geometry. The time step was 0.083 fs, box size - X=200$\lambda$ (propagation axis), Y=80$\lambda$ in 2D and X=200$\lambda$, Y=Z=24$\lambda$ in 3D, laser wavelength $\lambda$=0.8 $\mu$m, spatial resolution - $\lambda$/32 in 2D and $\lambda$/32 (along X) and $\lambda$/4 (along Y,Z) in 3D. Several transversal shifts 2.5-20 $\mu$m were simulated. The 50 fs pulse was focused in a 4 $\mu$m spot FWHM at the front surface of the non-perturbed tape with normalized vector potential $a_0$=1.5, where $a_0\approx0.85\sqrt{I\lambda^2}$, which corresponds to the experimentally available peak intensity of $\sim5\times10^{18}$ W cm$^{-2}$.

\begin{figure}[h]
\includegraphics[width=1\linewidth]{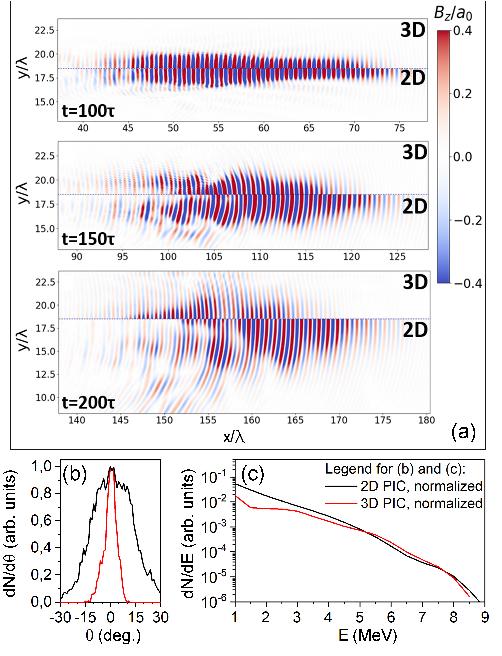}
\caption{\label{Figure4} Comparison of magnetic field of the laser pulse in 3D and 2D at different simulations periods (a). Normalized angular distribution of particles (b) and energy spectra (c).}
\end{figure}

Without focusing too much attention on the mechanisms of electron acceleration and interaction features (which can be found in \cite{25Ivan}), here we focus onto deflection of the femtosecond and electron beams while passing through the plasma slab with radial dependence of the electron density. It should be noted that this effect can be used not only to deflect a laser beam (as well as an electron beam \cite{23Mittelberger}), but also to diagnose the transverse plasma inhomogeneity itself \cite{30Brockington,31Enloe}. The principle difference between the 2D and 3D simulations occurs in the modeling of self-focusing \cite{32Pukhov} resulting in a longer lasting channel in the 3D case. Fig.\ref{Figure4}(a) demonstrates the compared evolution of the magnetic field $B_z$ in units of $a_0$ of the pulse in 3D/2D. The trailing part of the laser pulse diverges stronger in the 2D model during the propagation in the channel. The particles gain a finite energy in the channel within a few betatron oscillations and then go out of synchronism, still being trapped by the channel walls though. This affects the accelerated electron beam divergence: $\sim$0.5 rad in the 2D and $\sim$0.1 rad in the 3D, evaluated as the FWHM of the angular distribution of number of particles $\dfrac{dN}{d\theta}$, see Fig.\ref{Figure4}(b). Nevertheless, the spectra $\dfrac{dN}{dE}$ in these two cases, Fig.\ref{Figure4}(c), reasonably match in terms of shape (above 1 MeV) and maximal energy. Estimations of beam divergence in 3D correlates well with experimental measurements of electron beam.

\begin{figure}[h]
\includegraphics[width=1\linewidth]{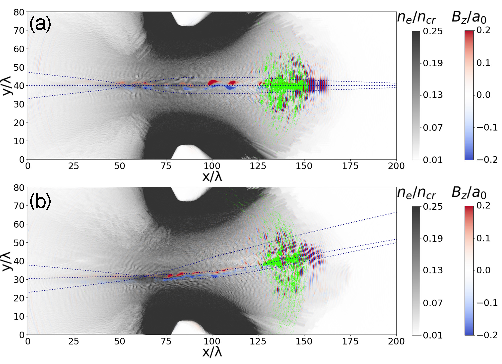}
\caption{\label{Figure5} Results of numerical 2D PIC calculations: normalized magnetic field in the box at the background of normalized electron density during propagation of a femtosecond pulse with deviation \textit{D} = 0 (a) and 10 $\mu$m (b). Green dots mark electrons with energies above 3 MeV. The blue dashed lines represent the results of ray tracing of a laser beam through an inhomogeneous plasma.}
\end{figure}

Fig.\ref{Figure5} represents a wider image of the beam propagation, ionization and particles motion in the simulation box (2D) with the axial and shifted (D = 10 $\mu$m) geometry. The concentration gradient promotes the beam deflection. As a result, the plasma channel also deviates from the original axis with conservation of its properties. It is important to note that the laser beam deflects mainly in the region of higher plasma density near the initial tape position, where the electron concentration $n_e$ after ionization lies in the range 0.1-0.2 units of critical density $n_{cr}$ (for 800 nm radiation $n_{cr}=1.7\times10^{21}$ cm$^{-3}$). This is supported by the ray-tracing through graded-index medium by solving the eikonal equation proposed by A.Sharma et al. \cite{33Sharma}: $\dfrac{d}{ds}n(\Vec{r})\dfrac{d}{ds}=\nabla{n(\Vec{r})}$, where $\Vec{r}$ is the position vector of the ray relative to origin, $ds$ is the element of the beam path, $n(\Vec{r})$ represents the refractive index distribution. The solution by 4th order Runge-Kutta method is shown by the dashed lines in Fig.\ref{Figure5}. It does not consider self-focusing but clearly depicts the refraction in the preformed plasma lens. The resulting effect in the bended channel is the simultaneous deflection of the accelerated electron bunch. 

\begin{figure}[t]
\includegraphics[width=1\linewidth]{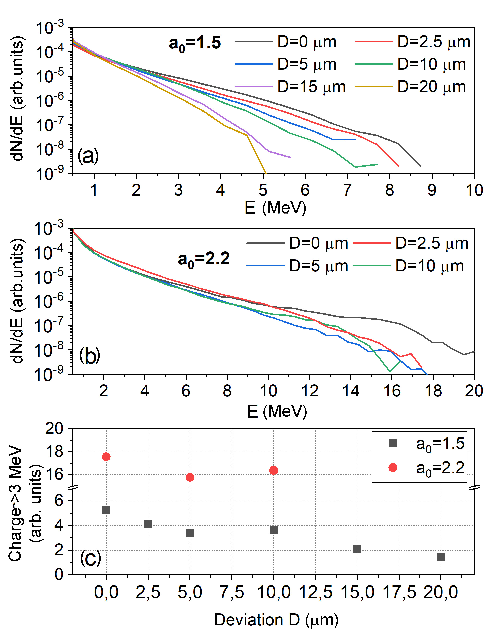}
\caption{\label{Figure6} Energy spectra of the electron beam at varied driver laser beam deviation \textit{D} from symmetry axis for the vector potential $a_0$=1.5 (a) and $a_0$=2.2 (b). The relative charge of the electron beam for particles above 3 MeV in dependence on the D (c).}
\end{figure}

The basic beam quality parameters are summarized in Fig.\ref{Figure6} in dependence on the deviation \textit{D} of the driver pulse for two values of $a_0$. The laser pulse begins to pass over the denser plasma region at \textit{D}$\textgreater$10 $\mu$m, where due to the modest peak intensity ($a_0$=1.5) the acceleration regime related to channeling and DLA struggles to evolve efficiently, which is seen as decrease of electrons energy, Fig.\ref{Figure6}(a). Our simulations with increased peak intensity ($a_0$=2.2, Fig.\ref{Figure6}(b)) demonstrate that this effect is less pronounced, ensuring experimental applicability of the proposed approach in a wide range of laser intensity and final electrons energies. One can also notice that charge of the bunch is almost the same independently of the deviation \textit{D} with $a_0$=2.2, Fig.\ref{Figure6}(c). As mentioned above the divergence in the 2D simulations is higher as compared to the 3D and the experiment. However the angular spread does not tend to widen in the investigated range of \textit{D} being in agreement with the experimental data.
A similar effect, considered earlier in the context of wakefield acceleration, led to a transverse inhomogeneity of the plasma wave, the accelerating longitudinal and focusing transverse fields of which deflected the accelerated electron beam following the laser driver \cite{21Nakajima,23Mittelberger}. The deviation of the components of the laser field from the initial direction itself determines the change in the trajectory of the accelerated electron. The effect of electron beam small angular deflection in the wake was also discussed in terms of transverse density gradient in gas jet outflow \cite{34Doss}. Recently the acceleration of electrons in the curved plasma channel was demonstrated experimentally \cite{35Zhu}.
Within the framework of the plasma model of an ablated and bored thin tape, one can conclude that key bunch parameters (charge, spectrum, divergence) are retained acceptable and comparable to the coaxial propagation in a wide range of deviations from the central axis ensuring beam deflection up to 10 degrees. When further increasing \textit{D} the laser pulse hits denser layers, and the acceleration regime becomes affected. Nevertheless, already small displacements evidently change the electron beam propagation axis by several degrees, which is more than enough for the beam deflection from its initial axis in whole range of particles energies without the use of energy selective magnetic optics.

\begin{figure}[h!]
\includegraphics[width=1\linewidth]{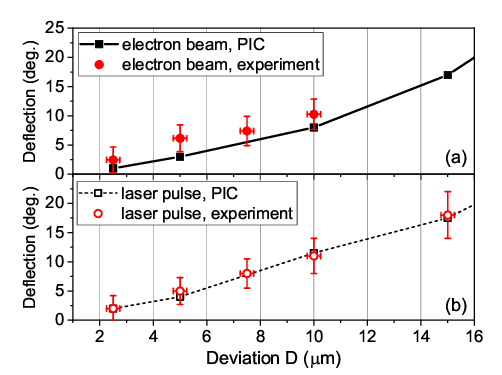}
\caption{\label{Figure7} Numerical and experimental data on the deflection of the electron beam (a) and the laser pulse (b) as a function of the femtosecond pulse deviation \textit{D} relative to the breakdown symmetry axis.}
\end{figure}

The summary of our studies with experimental data on the pulse and electron beam deflection (evaluated from scattered light signal and Lanex screen correspondingly) and comparison with numerical results are presented in Fig.\ref{Figure7}, where a good correlation is observed. This indicates that the numerical model of the tape target boring by the nanosecond prepulse is consistent with the experiment. Moreover, one can rely on the good evaluation of the electron bunch acceleration mechanism established numerically.

\section{Conclusion}

The angular manipulation of a laser-generated electron beam during its acceleration in a relativistic plasma channel in the presence of a radial electron density gradient is experimentally shown for the first time. An inhomogeneous transverse density distribution was formed under the action of a powerful nanosecond pulse ($10^{13}$ W cm$^{-2}$) on a thin mylar tape (15 $\mu$m). Heating and ablation lead to the tape boring with the formation of a plasma with a strong radial dependence of the concentration. The propagation of a relativistically intense femtosecond pulse (above $10^{18}$ W cm$^{-2}$, 50 fs) through a plasma with a slight deviation relative to the breakdown symmetry axis is accompanied by refraction of the laser beam on the density gradient.
It has been numerically established that, within the framework of the considered model of the breakdown and boring of a tape target by nanosecond radiation, the deviation of the propagation axis of a femtosecond pulse from the breakdown symmetry axis of up to 10 $\mu$m leads to  deflection of the generated electron beam by up to 10 degrees, while the beam parameters maintain almost unchanged. The calculations are confirmed by experimental data, where a similar deflection of the electron beam is observed on the detector.

\begin{figure}[h]
\includegraphics[width=0.8\linewidth]{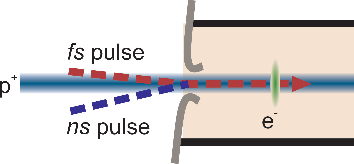}
\caption{\label{Figure8} Simplified model of the implementation of electron beam deflection approach to the AWAKE experiment.}
\end{figure}

The experimental approach can be used to create schemes for multistage electron acceleration in a sequence of laser accelerators. Of great importance is the possibility of in-situ control of beam directionality for the multistage accelerators, where the electron bunch needs to be injected at a small angle to the axis \cite{16Starodubyseva}. In the proposed approach there is no angular dispersion in the energy of particles in the beam, which preserves a high charge of the beam in a collimated form in contrast to the case of using magnetic optics. Our representation, for example, for the AWAKE experiment the scheme could be drawn as in Fig.\ref{Figure8}. The driver femtosecond pulse provides seed electrons with 10-20 MeV energy. The tape target may serve as the entrance window of the gas cell. It is also worth noting that the proposed method can serve not only for the tasks described above, but also for diagnosing the interaction region: an optically opaque volume of dense plasma near the tape, the channel formed by a nanosecond pulse, and the radial density distribution in it can be indirectly estimated from angle of refraction of the escaping beam.

\begin{acknowledgments}

The work was carried out with the support of the Russian Science Foundation project 21-79-10207. A part of equipment was purchased with the support of the Ministry of Education and Science of the Russian Federation within the framework of the national project "Science and universities". The applicability of our approach for multi stage electrons acceleration and AWAKE experiment was considered in the frame of scientific program of the National Center of Physics and Mathematics  (project ``Physics of high energy density. Stage 2023-2025'').

\end{acknowledgments}

\bibliographystyle{apsrev4-2}
\bibliography{references}

\end{document}